\documentclass{edp-conf}
\usepackage{graphicx}
\begin{document}

\TitreGlobal{SF2A 2002}

\title{Molecular gas in the double-barred Seyfert 2 galaxy NGC 5728} 
\author{Combes, F.}
\address{LERMA, Observatoire de Paris, FRANCE}
\author{Leon, S.}
\address{I. Physikalisches Institut, Universitaet zu Koeln, Germany}

\runningtitle{Molecular gas in NGC 5728}
\setcounter{page}{237}
\index{Combes, F.}
\index{Leon, S.}

\maketitle
\begin{abstract}We present CO(1-0) and CO(2-1)  maps of the Seyfert galaxy NGC 5728.
Although a stellar nuclear bar structure is clearly identified 
in the near-infrared images in the central
10", inside the ring identified as the ILR of the primary bar, there 
is no nuclear bar structure in the molecular gas.
Instead, the CO emission reveals an elongated structure,
15" in length, beginning at the nucleus (defined by the radio center)
aligned with the jet/ionisation cone, at a PA of 127 degrees.
This morphology, not frequently observed in Seyfert galaxies,
may be interpreted in terms of enhanced CO excitation
along the walls of the cone. Kinematical perturbations 
along the cone support this scenario. At larger-scale, CO 
emission is tracing the primary bar, and outer ring structure.
The total molecular mass, estimated from the CO emission, is 
M(H$_2$) = 3.1 10$^9$ M$_\odot$.\end{abstract}
%
\section{Introduction}

NGC 5728 is a well known prototype of embedded bars. It is classified
as (R1)SAB(r)a, and a Seyfert 2 (distance 33 Mpc -- 1''=180 pc). 
Its nuclear bar is oriented at 45 degrees of the main bar (Shaw et al 1993),
and could rotate faster.

Petitpas \& Wilson (2002) have obtained a CO map of NGC 5728 with the 
OVRO interferometer. They found a total CO mass
of 3.1 10$^8$ M$_\odot$. However, they underestimate the extended component, 
filtered out by the interferometer, and miss a significant part of the
CO emission, due to restricted velocity coverage.
We report here CO(1-0) and CO(2-1) observations with the IRAM-30m telescope,
with 23 and 12" beams.

\begin{figure}[h]
   \centering
   \includegraphics[width=11.5cm]{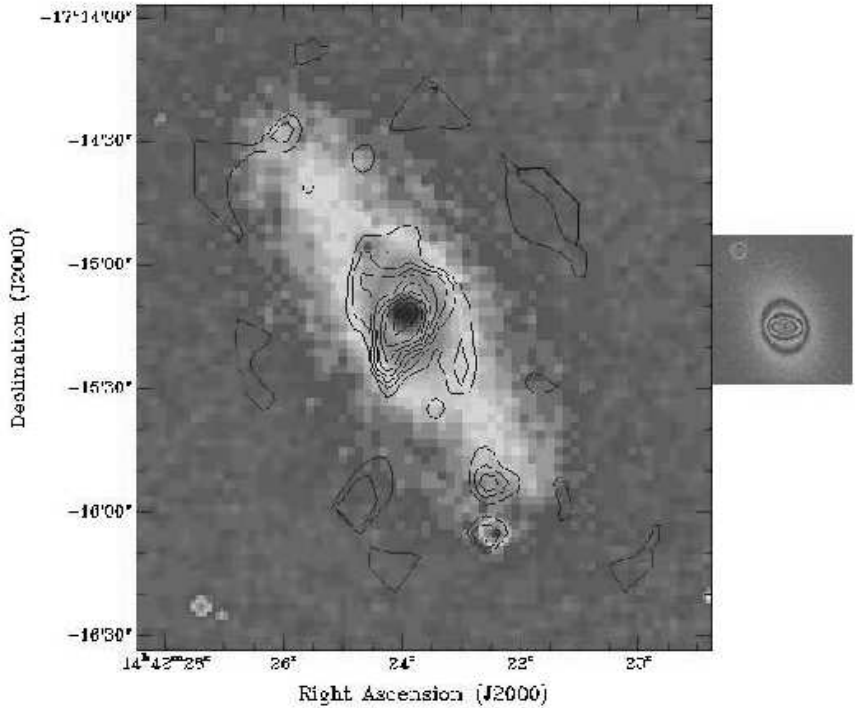}
      \caption{Contours of the CO(1-0) emission, superposed
on the 2-mass NIR image ({\it left}), and at the same scale
an adaptive optics NIR image (obtained with CFHT, Combes et al in prep)
showing the stellar nuclear bar ({\it right}).}
       \label{fig1}
   \end{figure}

\section{Main results}

The molecular gas distribution does not follow
the nuclear bar in the center, but is strongly asymmetric.
It follows the ionization cone, with the geometry found by Wilson et al (1993).
At larger scale, there is a tendency for the CO intensity
to follow the primary bar, along its characteristic dust lanes
(cf Fig. 1).

The total H$_2$ mass (without
 including the He) is  3.1 10$^9$ M$_\odot$ with a conversion factor of 3
10$^{20}$ cm-2/(K.km/s). It is larger than the HI mass of 2.5 10$^9$ M$_\odot$.
 At the tips of the bar, strong CO(1-0) emission is found, associated
with star formation, due to orbit crowding (in particular on the receding side,
at 35" $\sim$ 7kpc).
The total molecular gas mass is about 10 \% of the stellar disc mass.

The CO(2-1)/CO(1-0) intensity ratio is in general
quite low (between 0.5 and 1). One interesting feature is the presence of
two strips of higher CO ratio parallel to the radio jet direction. It could
indicate an enhancement of the CO  excitation along the wall of the ionization cone.

\end{document}